\documentstyle[preprint,aps,epsfig, multicol]{revtex}
\tighten
\draft
\widetext
\input{epsf.sty}
\preprint{}
\def\be{\begin{equation}}
\def\ee{\end{equation}}
\def\baray{\begin{eqnarray}}
\def\earay{\end{eqnarray}}
\def\ba{\begin{eqnarray}}
\def\ea{\end{eqnarray}}  
\def\V{{\bf V}}

\def\Dbar{\overline{D}}
\def\ap{{\alpha^{\prime}}}
\def\A{\hat A}
\def\Z{\hat Z}

\def\kp{k_\perp}
\def\kpl{k_\parallel}

\begin{document}

\title{Inter-Brane Potential and the Decay of a non-BPS-D-brane to 
Closed Strings}

\bigskip

\author{Saswat Sarangi \footnote{Electronic mail:
sash@mail.lepp.cornell.edu}
and S.-H. Henry Tye \footnote{Electronic mail:
tye@mail.lepp.cornell.edu}}

\address{Laboratory for Elementary Particle Physics, Cornell
  University, Ithaca, NY 14853}

\date{\today}

\bigskip

\maketitle

\begin{abstract}
We calculate the potential for $Dp-\Dbar p$ pair and show 
that the coincident $Dp-\Dbar p$ system has $(11-p)$ tachyonic 
modes, with $(9-p)$ of them due to radiative corrections.
We propose that the decay width of an unstable non-BPS-$Dp$-brane 
to closed strings is given by the imaginary part of the one-loop 
contribution to the effective potential of the open string 
tachyon mode. 
\end{abstract}

\section{Introduction}

The study of non-BPS brane systems are important in string theory.
Some of the dynamics involved play crucial roles in the 
inflationary scenario in the brane world \cite{rabadan,collection}.
We shall start by calculating the $Dp-\Dbar p$ potential per 
unit world volume $V(y)$, which is complex. We 
first examine the potential $V(\theta,y)$ between 
two $Dp$-branes at an angle $\theta$
and separation $y$. (The advantage of studying $V(\theta, y)$ 
first is clear : the underlying physics is much easier to 
keep track, since the open string spectrum, in particular 
the tachyon mode, depends on both $\theta$ and $y$. 
For $\theta=\pi$, $V(\pi,y)=V(y)$ is the potential 
between the $Dp-\Dbar p$ pair.) The lightest open string
mode has a mass that depends on $\theta$ and $y$. It is tachyonic 
for small $y$. The real part of the open string one-loop
contribution simply yields the closed string exchange potential
of the $Dp-\Dbar p$ pair separated at a distance $y$. The 
large $y$ behavior is simply Coulombic. However, at small $y$, 
this Coulombic behavior is truncated. Because the 
supersymmetry breaking is soft, a mass-level ``supersymmetric'' 
organization of the open string 
modes \cite{rabadan} yields a finite effective potential, as 
shown in Fig 1. For $V(y)$, the open string spectrum level crossing 
and the unbounded growth in the ``soft'' supersymmetry breaking
for massive modes is qualitatively different from the small 
$\theta$ case \cite{rabadan}. 
We shall see that the scalar modes ${\bf y}$ (there 
are $9-p$ of them) are also tachyonic when the $Dp$-brane is on 
top of the $\Dbar p$-brane. Their tachyonic property is due to 
quantum effects (radiative correction), a feature first 
seen in Ref\cite{rabadan} for small $\theta$.

\begin{figure}
\begin{center}
\epsfig{file=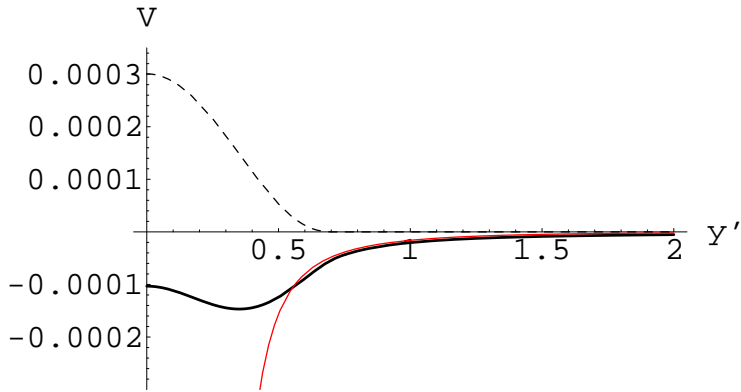, width=10cm}
\vspace{0.1in}
\caption{The potential $V(y)$ as a function of the separation $y$ 
for the $Dp-\Dbar p$-brane pair for $p=$4, where $\ap=$1. 
The dashed curve is the imaginary part of $V(y)$. 
The thick line is the real part of $V(y)$. The Coulombic 
potential (the thin red curve) is shown for comparison. 
}
\label{1}
\end{center}
\end{figure}
In the open string one-loop channel, since only the tachyon mode 
contribution has an imaginary part at the one-loop level, the 
evaluation of $Im~ V(y)$ is completely field theoretic \cite{wewu}. 
A simple generalization 
from 4 spacetime dimensions to $(p+1)$ dimensions yields the 
imaginary part of $V(y)$, as shown in Fig. 1.
\be
\label{imagVww}
Im V(y) = \frac{ \pi}{\Gamma ((p+3)/2)}
\left(\frac{ |m_{tachyon}^2|}{4 \pi} \right)^{(p+1)/2} 
\ee
where ($\ap$ is the open string Regge slope),
\be
\label{tachyonmass}
\ap m_{tachyon}^2 =  \frac{y^2}{4 \pi^2 \ap}  - \frac{1}{2}
\ee
Its physical meaning has been 
discussed extensively \cite{wewu,marcus,craps}. 

In the dual channel, that is, the closed string channel,
we should obtain exactly the same result. For large 
separation $y$, one obtains the well-known Coulombic potential
\cite{bachas}, due to the NS-NS and RR exchanges, which correspond 
to the attractive gravitational and RR forces. As we go to short 
distances, the massive closed string modes that 
are Yukawa-suppressed start contributing to the potential $V(y)$.
As $y \rightarrow 0$, we see that (naively) the potential 
diverges. This apparent divergence appears when the 
Hagedorn-like degeneracy overcomes the Yukawa suppression.
In fact, this happens precisely when the lightest 
open string mode becomes tachyonic. A regularization (that is, 
an analytic continuation of the integral) renders the result finite, 
but with an imaginary part, as expected, precisely reproducing the 
result (\ref{imagVww}) obtained in the open string channel. 
This is clearly related to the decay of the $Dp-\Dbar p$-brane pair.
When applied to a non-BPS $Dp$-brane, this same regularization 
approach yields a finite $Im~<Dp|$ closed strings $|Dp>$.

The decay of an unstable non-BPS D-brane
to closed strings has been studied extensively\cite{liu,gir,others}.
Adapting the above calculation to this case, we consider
$<Dp|~ \Delta~ |Dp>$ where $\Delta$ is the closed string propagator
and the sum over the closed string spectrum is implied.
This is equal to $V(0)/2$, which is finite with an imaginary part.
(The factor of $1/2$ is because the open string spectrum of the 
non-BPS $Dp$-brane is half of those that stretched between the
$Dp-\Dbar p$-brane pair at $y=0$).)
Using optical theorem (i.e., perturbative unitarity), as shown 
in Fig. 2, we interpret this as the decay width $\Gamma$ of the
non-BPS $Dp$-brane to on-shell bosonic closed string modes :
\be
\Gamma =  V_p~ Im~V(0) 
\ee
where $V_p$ is the $Dp$-brane world volume. As expected, $\Gamma$ 
is 0$th$ order in the string coupling. 
Here, the finite imaginary part appears due to the Hagedorn-like 
degeneracy of the massive closed string modes, and the analytic 
continuation moves the closed string modes from off-shell to on-shell. 
The decay first goes to very massive, non-relativistic on-shell 
bosonic closed string modes \cite{liu,gir}, with transverse 
momentum very small compared to the mass $m$ ($\kp/m \sim 1/\sqrt{m}$). 
These non-relativistic massive modes then decay to relativistic 
light closed string modes, both bosonic and fermionic.

\begin{figure}
\begin{center}
\epsfig{file=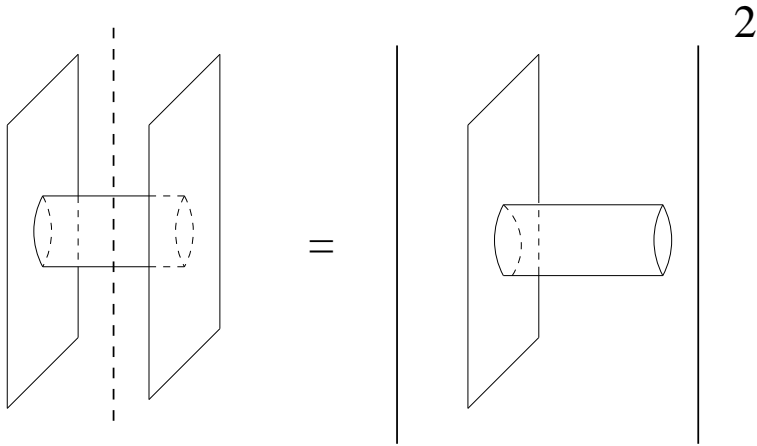, width=10cm}
\vspace{0.1in}
\caption{The vertical dashed line on the left side indicates 
taking the imaginary part of the $<Dp|$ closed strings $|Dp>$ 
amplitude in the decay of a non-BPS $Dp$ brane to closed string modes.
The right side is $|f(Dp \to$ closed string$)|^2$.}
\label{2}
\end{center}
\end{figure}

Intuitively, this is quite reasonable, since the imaginary part 
of the effective potential should be related to some decay width.
The above decay width is obtained perturbatively so it is expected 
to go to perturbative modes, i.e., closed and/or open string modes. 
In the decay of a non-BPS $Dp$-brane,
there is no brane left after the decay, so no open string modes 
can be present. Since only closed string modes are present at the 
end of the decay, the decay width $\Gamma$ should be a measure of 
the decay to closed string modes. (Lower dimensional branes are 
solitonic, which are non-perturbative, so their production during 
the decay should not be included in $\Gamma$.)   

Although $Im~V(y)$ is clearly related to the decay of the 
$Dp-\Dbar p$ pair, it is harder to interpret $-2V_p Im~V(y)$ 
as its decay to closed strings. However, since closed string modes 
are the only perturbative modes left after the annihilation of 
the $Dp-\Dbar p$ pair, its value may provide an estimate of the 
actual decay width to closed string modes. 
This result has applications to brane inflation 
\cite{collection,rabadan}.

We shall first obtain the $Dp-\Dbar p$ potential in the open 
string channel by extrapolating the potential for branes-at-angle.
Next we see how the same finite complex $Dp-\Dbar p$ potential 
emerges in the closed string channel. We then relate its 
imagainary part at zero separation to the decay of a 
non-BPS-$Dp$-brane to closed strings. 

\section{$D{p}-\Dbar {p}$ Potential}

Let us consider the potential $\V(y)$ per unit volume
between a parallel $Dp-\Dbar p$
pair separated by a distance $y$, where the $Dp$-branes are BPS with 
respect to each other. Let us write $\V(y) = 2 \tau_p  + V(y)$, 
where the $\tau_p$ is $Dp$-brane tension. We shall consider
$p \le 7$.
In the closed string channel, $V(y)$ is given by \cite{bachas,polchinski}
\baray
V(y) = -\int^{\infty}_{0}\frac{ds}{s}(\frac{16 \pi^{3}\ap} {s})^{-(p+1)/2} 
e^{-\frac{y^{2}}{s\ap}} (\frac{2\pi}{s})^{4}16 
\prod_{m=1}^{\infty}\frac{(1+w^{m})^{8}}{(1-w^{m})^{8}}
\label{closedV} 
\earay
where $w=e^{-s}$. At large $y$, the above integral is dominated 
by large $s$ (which corresponds to long cylinder), so
\be
V(y) \simeq - \frac{\kappa^2\tau_p^2}
{ \pi^{(9-p)/2}}
\Gamma((7-p)/2) \frac{1}{y^{7-p}} 
\ee
where $\kappa^2=8 \pi G_{10}$.
For $p<7$, $V(y)$ vanishes as $y \rightarrow \infty$.
$V(y) \sim \ln(y)$ for $p=7$.
This is simply the attractive NS-NS (gravitational) plus RR interaction
between the branes.

At short distances (small $s$), this expression naively diverges, although 
$V(y)$ is expected to approach a finite value even for $y=0$.
Let us first calculate $V(y)$ in the open string channel, which 
yields a finite $V(y)$ for all $y$. Then we come back and show 
how to obtain the same result in the closed string channel.  

\subsection{$Dp-\Dbar p$ Potential In the Open String Channel}

Using the Poisson resummation formula, the above $V(y)$ for the 
$Dp-\Dbar p$ system may be rewritten as the open string one-loop 
amplitude:
\baray
\label{openV}
&&V(y) = -\int^{\infty}_{0}\frac{dt}{t}(8 \pi^{2}\ap t)^{-(p+1)/2} 
e^{-t y^{2}/2\pi\ap} Z(\pi,t) \\\nonumber
&&Z(\pi, t) = \frac{1}{q^{1/2}}\prod_{m=1}^{\infty}
\frac{(1-q^{m-1/2})^8}{(1-q^{m})^8} \\\nonumber
&&= \frac{1}{q^{1/2}} - 8 + 36 q^{1/2} -128q + 402 q^{3/2}
- 1152 q^{2}+3064 q^{5/2}- 7680q^{3}\nonumber \\ 
&&+ 18351 q^{7/2}- 42112 q^{4} + 93300 q^{9/2}- 200448 q^{5}+ ...
\nonumber
\earay
where $q=e^{-2 \pi t}$.
This sum is over the open string modes. For large $y$, this sum 
converges rapidly to give $V(y)$. Notice that as $y$ decreases,
the mass of the lightest open string mode is given in 
(\ref{tachyonmass}),
which becomes tachyonic for small values of $y$. In the presence 
of the tachyon, the above sum apparently 
becomes ill-defined, that is, the integral diverges.
The tachyonic mode contribution to $V(y)$ is 
\baray
\label{openVt}
&&V(y) = -\int^{\infty}_{0}\frac{dt}{t}(8 \pi^{2}\ap t)^{-(p+1)/2}
\exp \left( - 2 \pi t (y^{2}/4\pi^2\ap - 1/2)\right) 
\earay
which is divergent. This integral can be regularized by analytic 
continuation \cite{marcus}. 

%\begin{figure}
%\begin{center}
%\epsfig{file=contour.eps, width=10cm}
%\vspace{0.1in}
%\caption{The contour of integration.}
%\label{3}
%\end{center}
%\end{figure}

%We use the result for the contour as shown,
%\baray
%\int_{C}\frac{dx}{x}x^{-\alpha}e^{x}= i\frac{2\pi}{\Gamma(1+\alpha)}
%\nonumber
%\earay
%To extract the imaginary part we are looking for, take the part of 
%the contour that spans from $+0$ to $-\infty$ in the lower half of 
%the complex plane and notice that it is the same as the integral 
%with limits from $+0$ to $+\infty$ (the dotted line in the lower 
%half plane). This can be seen easily by noting that the integral 
%vanishes for a closed semicircular contour (counterclockwise) that 
%encloses the lower half complex plane with the real axis as the 
%diameter of the semicircular contour. 

To see how the divergence can be regularized using analytic 
continuation, consider a slightly different integral:
\baray
 \int_{\delta}^{\infty}\frac{dx}{x}x^{- \alpha}e^{(a+ i\epsilon)x} = 
\left( a \exp[-i(\pi - \epsilon)]  \right)^{\alpha} 
\int_{-a\delta - i\epsilon}^{\infty} \frac{dx}{x}~x^{-\alpha}~ e^{-x}
\nonumber
\earay
This integral has a branch cut along the negative $x$ axis. The 
$i \epsilon$ prescription tells us to integrate under the branch cut. 
The integral is now finite. The divergence has been removed in favor 
of an imaginary part which does not depend on the value of $\delta$. 
The divergence is not an infinity of the theory, but rather an 
indication of the amplitude becoming complex. The result of the 
analytic continuation is :
\baray
\label{imagin}
Im \lgroup \int_{0}^{\infty}\frac{dx}{x}x^{- \alpha}e^{ax}\rgroup
= \frac{\pi}{\Gamma(1+\alpha)} a^{\alpha}  
\earay
for positive $a$. Using (\ref{imagin}), this gives, for 
$y^2 < 2 \pi^2 \ap $, the imaginary part of $V(y)$ : 
\be
\label{imaV}
 Im ~  V(y) = \frac{\pi}{\Gamma ((p+3)/2)}
\left(\frac{ |m_{tachyon}^2|}{4 \pi} \right)^{(p+1)/2}
\ee
This $Im ~  V(y)$ is shown in Fig. 1.
Since only the tachyon mode contributes to the imaginary part 
of $V(y)$, we can also evaluate $Im ~  V(y)$ using the standard 
quantum field theory method. 
Its contribution can be calculated via the Coleman-Weinberg 
effective potential \cite{wewu} where
the one loop vacuum-to-vacuum amplitude for a point particle of 
mass m in $(p+1)$ dimensions is given by:
\baray
Z(m^{2}) = iV_{p+1}\int_{0}^{\infty}\frac{dl}{2l}
\int\frac{d^{p+1}k}{(2\pi)^{p+1}}e^{-(k^{2}+m^{2})l/2} 
= \frac{iV_{p+1}}{(2\pi)^{p+1}} \int_{0}^{\infty}\frac{dl}{2l}
\frac{e^{-m^{2}l/2}}{l^{(p+1)/2}}
\earay
Inserting the open string tachyon mass $\ap m^{2} <0$ we see that 
the above integral diverges. After a proper regularization using 
analytic continuation, we get the imaginary part of the energy 
density Im(E) as (\ref{imaV}).

To calculate the real part of $V(y)$ one has to include the whole tower 
of open string modes. The result depends on how 
the oscillating terms are grouped. The particular way of grouping 
the terms should be dictated by the soft supersymmetry breaking,
as suggested by Garcia-Bellido, Rabadan and Zamora \cite{rabadan}.
They applied it to the branes-at-small-angle case, where $\theta=$0 
corresponds to two parallel BPD $Dp$-branes while $\theta=\pi$ 
corresponds to the $Dp-\Dbar p$-brane pair. For the 
$Dp-\Dbar p$ system, supersymmetry breaking becomes large and 
level-crossings happen. We  show that, despite these, the 
convergence remains intact. 

When the two branes are parallel there is no potential between 
them because of the supersymmetry. Each mass level contains a 
set of the supermultiplets. The contribution to the potential 
$V(y)$ from the open string bosons is exactly cancelled by the 
contribution from the open string fermions, mass level by mass level.
We keep this mass level grouping. The way we identify the grouping 
as the angle $\theta$ between the branes increases from zero to $\pi$
is by following the spectral flow, that is, the splitting of open 
string modes at each level due to the soft supersymmetric breaking. 
This splitting leads to unequal contributions from the bosons and 
the fermions, resulting in a finite potential between the brane 
and the antibrane.

\subsection{Branes-at-an-Angle Potential}      

Let us first review the branes-at-an-angle case. For simplicity, 
let us consider the case of two $D$4-branes at an angle $\theta$.

The potential for two $D$4-branes at an 
angle $\theta$ in the open string one-loop channel is given by, 
\baray
V(y,\theta) = -V_{4}\int^{\infty}_{0}\frac{dt}{t}(8 \pi^{2}
\alpha^{\prime}t)^{-2} e^{- \frac{t}{2\pi\alpha^{\prime}}
y^{2}}Z(\theta,t),\nonumber \\
Z(\theta,t) = \frac{\Theta_{11}^{4}(i\theta t/2\pi, it)}
{i\Theta_{11}(i\theta t/\pi, it)\eta^{9}(it)}
\earay
For $\theta=$0, the $Dp$-brane pair is supersymmetric and 
$Z(\theta = 0,t)=$0.

\begin{figure}
\begin{center}
\epsfig{file=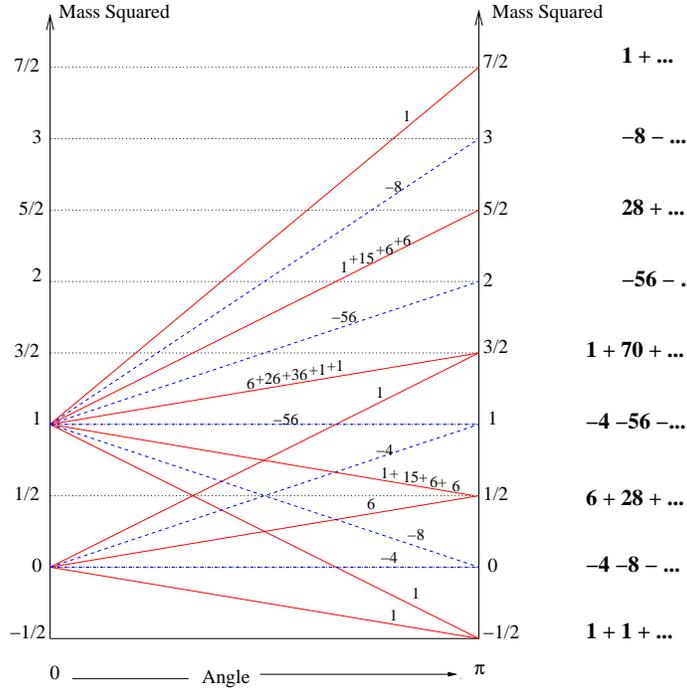, width=9cm}
\vspace{0.1in}
\caption{The spectral flow of the lowest two level open string 
modes as a function of the angle $\theta$ between the branes for 
$y=0$. $\theta=$0 corresponds to the 2 parallel BPS $Dp$-branes
(left) while $\theta=\pi$ corresponds to the $Dp-\Dbar p$ case 
(right). The red (solid) lines show the NS states, the blue 
(dashed) lines show the R states. The number written above a 
solid/dashed line is the number of states represented by the line.
The numbers in boldface to the right of the diagram show the 
states at the given mass levels for the $\theta = \pi$ case. 
$+$ is for bosons and $-$ is for fermions.}
\label{4}
\end{center}
\end{figure}

The splitting of states is shown in the figure 3. To calculate the
splitting one observes that the $NS$ sector zero point energy has
an angle dependence given by $\epsilon_{NS}= -1/2 + \theta /2\pi$.
The R sector zero point energy remains unaffected by the angle 
between the branes. Considering $D4$ branes in the light cone gauge, 
we take the branes to make an angle $\theta$ in the $7-8$ plane,
then the creation operators in these two dimensions are given by
$b_{-r + \theta /\pi}^{7,8}$ ,where $r = 1/2, 3/2, ...$ , and for
the rest of the six dimensions the creation operators are $b_{-r}^{i}$,
where $i= 1,2,...,6$.  Similarly the creation operators in the $R$
sector are given by $d_{-n + \theta / 2\pi}^{7,8}$ for the $7$ and
the $8$ dimension and by $d_{-n}^{i}$ for the other six dimensions,
where $n = 0,1,2,...$. Note that since we have only six zero modes
for the $R$ sector for nonzero $\theta$, therefore, the fermions
will be in $4$-dimensional representation of the Dirac algebra.
The above consideration suggest the following grouping of the 
terms in the partition function:
\baray
Z(t)= \frac{\Theta_{11}(it/2, it)^{4}}{e^{\pi t}\eta(it)^{12}} 
= \frac{(q^{1/2}-1)^{4}}{q^{1/2}}\prod_{m=1}^{\infty}
\frac{(1-q^{m+1/2})^{4}(1-q^{m-1/2})^{4}} {(1-q^{m})^{8}}
\earay

The long distance behavior is determined by the $t \to 0$ 
limit of the partition function. In this limit, 
$Z(\theta,t) \to 4t^{3}\sin^{2}(\theta /2)\tan(\theta /2)$, and 
the potential becomes:
\baray
V(y,\theta) = -\frac{\sin^{2}(\theta /2)\tan(\theta /2)}{8\pi^{3}
\alpha^{\prime}y^{2}}
\earay

The short distance behavior is determined by the 
$t \to \infty$ limit. The open string spectrum splits into copies 
of the (broken) supermultiplet. The SUSY breaking is due to 
the ``expectation value'' $\theta$, which is a soft SUSY breaking,
so that, for any $y$,
\baray
\sum_{i}(-1)^{F}m_{i}^{2n}=0, \quad \quad n=1,2,3
\earay
where $i$ runs over the spectrum in each ``softly broken'' supermultiplet 
(corresponding to a $N=4$ supermultiplet). 
The lightest open string mode has mass
\be
\ap m^2_{tachyon}= \frac{y^2}{4 \pi^2 \ap} -\frac{\theta}{2\pi}
\ee
In Fig 3, the open string 
modes are grouped into integer mass levels at $\theta=$0, 
each of which contains a set of (broken) supermultiplets.
Following the spectral flow and keeping this grouping for 
non-zero $\theta$, $Z(\theta, t)$ can be written as :
\baray
Z(\theta, t) &=& \frac{(z-1)^{4}}{z}
(\sum_{n=0}^{\infty}z^{2n})\prod_{m=1}^{\infty}\frac{(1-q^{m}z)^{4}
(1-q^{m}z^{-1})^{4}} {(1-q^{m})^{6}(1-q^{m}z^{2})(1-q^{m}z^{-2})}\\\nonumber
&=& \frac{(z-1)^{4}}{z}(\sum_{n=0}^{\infty}z^{2n})
[1 + \frac{(z-1)^{4}}{z^{2}}q + \frac{(z-1)^{4}(1+7z^{2}+z^{4})}
{z^{4}}q^{2}+ ...]
\earay
where $q=e^{-2\pi t}$ and $z=e^{-\theta t}$. Here, for each mass level,
the sum of the Landau levels in $\sum z^{2n}$ ....
After the integration over $t$, we obtain
\baray
V_{1-loop}= \frac{-1}{(8\pi \alpha^{\prime})^{2}}
\int_{0}^{\infty}\frac{dt}{t^{3}} \sum_{i}(-1)^{F_{i}}
e^{-2\pi \alpha^{\prime}tm_{i}^{2}} \\\nonumber
=\frac{1}{32\pi^{2}}\sum_{i}(-1)^{F_{i}}m_{i}^{4}
\ln(2\pi\alpha^{\prime}m_{i}^{2})
\earay
Notice that $V_{1-loop}$ is finite, since the quadratic divergence is 
absent for each multiplet.

Now let us go back to the $Dp$-$\Dbar p$ case.
The above expression diverges in this limit. This divergence is well 
understood: it is due to the appearance of an additional dimension 
in the overlapping world volume of the $Dp$-branes. To remove this 
divergence, we make the substitution:
\baray
\Theta_{11}(i\theta t/\pi, it) \to iL(8 \pi^{2}\alpha^{\prime}t)^{-1/2}
e^{-\pi t}\eta(it)^{-3}
\earay
This substitution gives us the expected partition function for 
the brane-antibrane case:
\baray
Z(\theta, t) \to Z(\pi, t) = (q)^{1/2}\frac{\Theta_{11}(it/2, it)^{4}}
{\eta(it)^{12}}\\ \nonumber
= \frac{1}{(q)^{1/2}}\prod_{m=1}^{\infty}\frac{(1-q^{m-1/2})^{8}}
{(1-q^{m})^{8}}\earay
This gives the $Dp$-$\Dbar p$ potential $V(y)$ in the open string 
channel (\ref{openV}).

To be specific, let us now focus on the $p=4$ case.
The long distance behavior is found 
by taking the $t \to 0$ limit of the partition function, with  
$Z(\pi, t) \longrightarrow 16 t^{4}$, we have 
\baray
V(y, \pi) = -2.56 \times 10^{-4}(\frac{2\pi \alpha^{\prime}}{y^{2}})^{3/2} 
\earay

To see the short distance behavior, retain the mutiplet grouping
structure at $\theta = \pi$, we write the partition function as 
a function of $q$ and $z=\sqrt{q}$ :
\baray
&& Z(\pi, t)= \frac{(z-1)^{4}}{z}\prod_{m=1}^{\infty}
\frac{(1-q^{m}z)^{4}(1-q^{m}z^{-1})^{4}} {(1-q^{m})^{8}} \\\nonumber
&&= \frac{(z-1)^{4}}{z}[1 -4 \frac{(z-1)^{2}}{z}q +
6 \frac{(z-1)^{2}(1-4z+z^{2})}{z^{2}}q^{2}+ ...]
\earay
We use this to evaluate the following integral which we shall need:
\baray
V_{1-loop}= \frac{-1}{(8\pi \alpha^{\prime})^{5/2}}\int_{0}^{\infty}
\frac{dt}{t^{7/2}} \sum_{i}(-1)^{F_{i}}e^{-2\pi \ap tm_{i}^{2}}\\\nonumber
=\frac{1}{(8\pi \alpha^{\prime})^{5/2}}\sum_{i}(-1)^{F_{i}}
\frac{8\Gamma(1/2)}{15}(2\pi\alpha^{\prime}m_{i}^{2})^{5/2}
\earay
where we have used the soft SUSY breaking condition to evaluate this 
integral. 
Now we can write down the potential terms for different 
string levels (taking the relevant power of $q$ from the above 
expansion :

Order $q^{0}$ in $V(y, \pi)$:
\baray
\frac{8\Gamma(1/2)}{15(4\pi )^{5/2}}\lbrack 
(y^{\prime2}- \theta^{\prime})^{5/2} - 4(y^{\prime2})^{5/2} + 
6(y^{\prime2}+ \theta^{\prime})^{5/2}-4(y^{\prime2}+
2\theta^{\prime})^{5/2} + (y^{\prime2}+ 3\theta^{\prime})^{5/2}\rbrack
\earay

Order $q^{1}$ in $V(y, \pi)$:
\baray
\frac{8\Gamma(1/2)}{15(4\pi )^{5/2}}\lbrack -4(1+ y^{\prime2}- 2\theta^{\prime})^{5/2} +24(1+ y^{\prime2}- \theta^{\prime})^{5/2} -60(1+ y^{\prime2})^{5/2} + 80\\ \nonumber(1+ y^{\prime2}+ \theta^{\prime})^{5/2} -60(1 + y^{\prime2}+2\theta^{\prime})^{5/2} + 24(1 + y^{\prime2}+ 3\theta^{\prime})^{5/2}-4(1+ y^{\prime2}+4\theta^{\prime})^{5/2}\rbrack
\earay
Order $q^{2}$ in $V(y, \pi)$:
\baray
\frac{6 \times 8\Gamma(1/2)}{15(4\pi )^{5/2}}
\lbrack(2+ y^{\prime2}- 3\theta^{\prime})^{5/2} -10(2+ 
y^{\prime2}- 2\theta^{\prime})^{5/2} +40(2+ y^{\prime2}- 
\theta^{\prime})^{5/2}\\ \nonumber 
-86(2+ y^{\prime2})^{5/2}  + 110(2+ y^{\prime2}+ 
\theta^{\prime})^{5/2} -86(2 + y^{\prime2}+2\theta^{\prime})^{5/2} + 
40(2 + y^{\prime2}+ 3\theta^{\prime})^{5/2}\\ \nonumber
-10(2+ y^{\prime2}+4\theta^{\prime})^{5/2}+
(2+ y^{\prime2}+5\theta^{\prime})^{5/2} \rbrack
\earay
and so on, where $y^{\prime}$ stands for $y/(2\pi\alpha^{\prime})$ and 
$\theta^{\prime}$ stands for $\theta/(2\pi\alpha^{\prime})$.
This yields the potential in Fig. 1. The imaginary part comes only 
from the lightest open string mode when it becomes tachyonic. The 
real part converges quite rapidly, we need to keep only up to $q^3$
to get within 1$\%$ accuracy.

The behavior of the potential for short distances can be seen 
from Fig. 1. An interesting result first observed in Ref\cite{rabadan}
is the emergence of new tachyonic modes in the ${\bf y}$-directions. 
There is a dip in the potential close to the origin and the second 
derivative of the potential at the origin is negative. This yields a 
tachyonic mass for ${\bf y}$ as a function of $\theta$.
\baray
\ap m^2(y) &=& 3.36 \times 10^{-4} \theta/\pi \quad 
\quad \theta \ne \pi \\ \nonumber
\ap m^2(y) &=& 4.51 \times 10^{-5} \quad \quad \theta = \pi
\earay

The appearance of the ${\bf y}$ tachyons happens for generic $p$.
In the open string classical limit, the brane separation ${\bf y}$ 
are $(9-p)$ moduli, so this tachyonic mass is a consequence 
of the one-loop open string 
contribution; that is, radiative corrections analogous to the 
Coleman-Weinberg mechanism. 
This feature first appears in the branes at small $\theta$ system
\cite{rabadan}. We see that it persists for all values of $\theta>0$.
Since this ${\bf y}$ tachyon mass is much smaller than 
the lightest open string tachyonic mode, we expect the latter to
dominate the brane dynamics at short distances.

\section{The $D{p}-\Dbar {p}$ Potential in the Closed String Channel}

The next step is to evaluate the same brane-antibrane potential 
$V(y)$ from the closed-string perspective.
The $Dp$-brane is a solitonic object and it can emit/absorb
closed string modes with arbitrary transverse momentum ${\bf \kp}$. 
\baray
V(y) = -\sum_j  \frac{1}{2^6 \pi}(4 \pi^2 \ap)^{4-p}
\int^{\infty}_{0}ds
\int \frac{d^{9-p}\kp}{(2\pi)^{9-p}}
\exp \left( -s \ap (\kp^2 + m^2_j)/4 
+ i {\bf \kp}\cdot {\bf y} \right) \label{closedVz} \\
= -\int^{\infty}_{0}\frac{ds}{s}(\frac{16 \pi^{3}
\alpha^{\prime}}{s})^{-(p+1)/2} e^{- {y^{2}}/({s\alpha^{\prime}})}
\left(\frac{2\pi}{s}\right)^4 Z(s) \\ \nonumber
\earay
\baray
Z(s) = 16 \prod_{m=1}^{\infty}\frac{(1+w^{m})^{8}}{(1-w^{m})^{8}} 
=  \sum_{0}^{\infty}A(n)w^{n}
\earay
where $n=\ap m^2/4$ and $w=e^{-s}$. This is (\ref{closedV}),
where 
\baray
\label{Acontour}
A(n)= \frac{1}{2\pi i} \oint dw \frac{Z(w)}{w^{n+1}} = \oint dz e^{nz} Z(s)
\earay
We see that only bosonic closed string (NS-NS and RR) modes
are included. Long distance behavior is governed by the light 
($n=0$) closed string modes 
which dominate the s-channel contribution to the 
potential in the $s \to \infty$ limit.
The result is finite and goes as $-y^{p-7}$.

Next, we consider the small $y$ (small $s$ and large $n$) behavior. 
Notice that $A(n)$ grows monotonically with $n$. So,
unlike the open-string calculation, where the coefficients in the
expansion for $Z(s)$ oscillate in sign so that they can be grouped
in a multiplet structure leading to a convergent sum, here the sum
grows monotonically and will lead to a divergence if the degeneracy 
factor dominates the Yukawa suppressing factor $ \exp({- y^{2}/s\ap})$.
To get an idea of the behavior, we need the large $n$ behavior of 
$A(n)$. As $s \to 0$ (see Appendix):
\baray
\label{zas}
Z(s) \simeq \left(\frac{s}{2\pi}\right)^4
\exp(\frac{2\pi^{2}}{s})
\earay
and $A(n)$ is obtained via the saddle-point approximation with 
the asymptotic form of $Z(s)$ (\ref{zas}) in (\ref{Acontour}), 
\baray
\label{an} 
A(n) \to  (2n)^{-11/4} e^{(8\pi^{2}n)^{1/2}}
\left( 1 + O(\frac{1}{\sqrt{2 n}}) \right)
\earay
Next, we perform the s-integration in $V(y)$ to obtain, in the 
large $n$ approximation :
\baray
V(y) &\sim& - y^{p-7} \sum_n A(n) e^{-2 y\sqrt{n \ap}} 
(1 + O(y\sqrt{n/\ap}) + ..) \\\nonumber
 &\sim & - y^{p-7} \sum_{n} n^{-11/4}  \exp\{4\pi \sqrt{n}(1/\sqrt{2}-
y/2 \pi \sqrt{\ap})\} \nonumber
\earay
the mass of the closed string mode being exchanged is
$2 \sqrt{n \ap}$ so that, for large values 
of $y$, the Yukawa suppression factor
$\exp (-2\sqrt{n}y/\sqrt{\ap})$ leads to a finite
potential. However, for $y\leq \sqrt{2 \ap} \pi $, 
the exponent blows up for large n, so the potential $V(y)$
apparently diverges. That is, the exponentially 
large (Hagedorn-like) number of massive closed string modes 
contributing to the potential overcomes the Yukawa suppression.
Comparing with the open string tachyon mass formula 
(\ref{tachyonmass}), we see that the divergence appears 
exactly at the value of $y$ where the lightest open string 
mode becomes tachyonic.  To emphasize this point,
let us go back to $V(y)$ (\ref{closedV}). In the large $n$
approximation, we replace $A(n)$ by its asymptotic value and 
replace the sum over $n$ by an integration over a continuous 
$u$ variable, where $2n=u^2$ and $s=2\pi/t$ :
\baray
\label{closedVn}
V(y) &\simeq& -\int^{\infty}_{0}\frac{dt}{t} \sum_n
(8 \pi^{2}\ap t)^{-(p+1)/2} t^{4}(2n)^{-11/4}
\exp\{2 \pi \sqrt{2n} - t \frac{y^{2}}{2\pi\ap} - \frac{2\pi n}{t}\}
\\\nonumber
&\simeq& -\int^{\infty}_{0}\frac{dt}{t} \int^{\infty}_{0} \frac{du}{u} 
(8 \pi^{2}\ap t)^{-(p+1)/2} t^{4} u^{-7/2}
\exp\{-\pi(u-t)^2/t  - 2 \pi \ap m^2_{t} t\}
\earay
where $\ap m^2_{t}=y^2/4 \pi^2 \ap  - 1/2$ is precisely
the mass squared of the lightest open string mode given in 
(\ref{tachyonmass}).
The integral over $u$ is gaussian-like so it is convergent,
while the integral over $t$ depends on $y$. For small $y$, 
$m^2_{t}<0 $ is tachyonic and the integral over $t$ diverges.

Now we are ready to get $Im~ V(y)$ in the closed string channel. 
Since the open string channel calculation of
the potential is finite everywhere, we expect the same for the 
closed string calculation of $V(y)$.
The appearance of this divergence at small $y$ signals the 
appearance of an imaginary part of the potential. 
We may use Eq(\ref{imagin}) to obtain the imaginary part of 
$V(y)$ in (\ref{closedVn}) when $m^2_{t} <0$. This gives a crude 
value of $Im~ V(y)$. It turns out 
that the correction (i.e., subleading) 
terms in $A(n)$ are important in the evaluation of $Im~ V(y)$,
and the convergence seems slow.

In fact, it is easier to evaluate $Im~ V(y)$ simply by going back
to $Z(s)$. Putting the asymptotic ($s \rightarrow 0$) 
form of $Z(s)$ (\ref{zas}) into $V(y)$ (\ref{closedV}) and 
using (\ref{imagin}), we obtain precisely the $Im ~V(y)$ given 
in (\ref{imaV}).  
As we shall see, corrections to the $s \rightarrow 0$
form of $Z(s)$ do not contribute to $Im ~V(y)$.
While the imaginary part seen in the open string channel 
is due to the single tachyonic mode, the imaginary part 
in the closed string channel is due to the cumulative effect
of the Hagedorn-like degeneracy of the massive closed 
string states. These two effects are dual to each other.
This point will become even more explicit below.

We still have to see how the real part of the 
potential $Re~ V(y)$ emerges in the closed string channel.
This is carried out by by an exact correspondence between 
the open string calculation and the closed string calculation.
To demonstrate this exact correspondence, we shall start from the closed
string channel and show that taking into account all orders
of correction to the asymptotic expression for $Z(s)$ leads
back to the open string channel expression. Furthermore, the
asymptotic expression for $Z(s)$ (\ref{zas}) corresponds 
precisely to the open string tachyon mode.
This shall also answer the question as to what happens to the
supersymmetric grouping of the terms present in the open
string partition function (that we make use of through the
soft SUSY breaking condition to find a finite 
sum in the open string channel) when we go
to the closed string channel.

We follow the calculation of the asymptotic form for $Z(s)$
making use of the Hardy-Ramanujan formula | Appendix| but
this time by keeping track of the correction terms that
would arise in the $s \neq 0$ region. The result is:
\baray
Z(s) = 16 e^{2\pi^{2}/s}(\frac{s}{4\pi})^{4} e^{\chi} \nonumber
\earay
where $\chi$ contains the correction terms,
\baray
\chi = -De^{-2\pi^{2}/s}+ \frac{D}{2}e^{-4\pi^{2}/s}-\frac{D}{3}
e^{-6\pi^{2}/s}+\frac{5D}{4}e^{-8\pi^{2}/s} + .....
\earay
Expanding the $e^{\chi}$ term for $D=8$, we get:
\baray
Z(s) = (\frac{s}{2\pi})^{4}\lgroup e^{\frac{2\pi^{2}}{s}} -8 +
36e^{-\frac{2\pi^{2}}{s}}-128e^{\frac{-4\pi^{2}}{s}}+
402e^{-6\frac{\pi^{2}}{s}}- ... \rgroup
\earay
Substituting this into the closed string channel $V(y)$ 
(\ref{closedV}), and making a change of variable
$s \to \frac{2\pi}{t}$, we recover the potential $V(y)$ in the open 
string channel (\ref{openV}):
\baray
V(y) = -\int_{0}^{\infty}\frac{dt}{t}(8\pi^{2}\ap t)^{-5/2}
e^{\frac{-y^{2}t}{2\pi\alpha^{\prime}}} \lgroup e^{\pi t} -8 +
36e^{-\pi t}-128e^{-2\pi t}+ 402e^{-3 \pi t}- ... \rgroup
\earay

One clearly sees that the imaginary part of the potential 
using the asymptotic form of $Z(s)$ in the closed string channel 
is exactly equivalent to the tachyon contribution in the open 
string channel. The massless and massive level contributions 
to the potential in the open string channel are exactly equal 
to the contribution coming from the corrections to this asymptotic 
form. They do not contribute to $Im ~V(y)$. One
could go on and start calculating the real part of the potential in
the closed string channel by taking the term by term contribution
coming from the subleading terms to $Z(s)$, grouping them as 
suggested by the soft SUSY breaking in the open string channel.

\section{Decay of A non-BPS Brane to Closed Strings}

The decay rate per unit world volume of a non BPS $Dp$-brane to 
closed strings can be written as
the square of the amplitude $|f(Dp \to closed~ string)|^2$. 
Via the optical theorem, we expect this to be given by 
$Im ~ <Dp|~\Delta~ |Dp>$, as shown in Fig. 2. $\Delta$ is the 
closed string propagator given by 
$ \int_{0}^{\infty}ds~e^{-s(L_{0}+ \tilde{L}_{0})/2}$, 
$L_{0}$ and $\tilde{L}_{0}$ being the Virasoro generators, and
$(L_{0}-\tilde{L}_{0})|Dp>=0$. Let us consider 
an unstable non-BPS-$Dp$-brane in Type II string theory.

For a non-BPS-$Dp$-brane, both ends of an open string end in the 
same brane, so $y=0$. For the same reason,
the open string spectrum in a non-BPS-$Dp$-brane has only one 
real tachyon mode, instead 
of a complex tachyon field, and it has only half of the spectrum of 
open strings that stretch between the branes in the $Dp$-$\Dbar p$
system at $y=0$. So the analysis of the one-loop open string 
effective potential in this case is identical to that
for $V(0)/2$, and we obtain $Im~V(0)/2$ as given in 
(\ref{imaV}) with $\ap m_t^2=-1/2$. 
Performing the $s$-integration in (\ref{closedVz}) at $y=0$,
we obtain
\baray
\label{Vprop}
<Dp|~\Delta~ |Dp> =   \frac{V(0)}{2}  =
-\sum_j \int \frac{d^{9-p}\kp}{(2\pi)^{9-p}}
\frac{\pi}{8}\frac{(4 \pi^2 \ap)^{3-p}}
{\kp^2 +  m^2_j + i \epsilon}
\earay 
where the sum is over closed string modes. Recall that
\baray
A(Dp~ \to~ Dp) \simeq  \sum_j \frac{f_j(Dp~\to~X_j) f_j(X_j~ \to~Dp)}
{k^{2} - m_j^{2} + i\epsilon} 
\to \sum_j i\delta(k^2-m_j^2) |f_j|^2
\label{factor}
\earay 
where, ignoring momenta $\kpl$ parallel to the brane, $k^2=k_0^2 -\kp^2$.
We see that $|f_j|$ is a constant independent of the closed string 
mode $j$, and that the imaginary part comes only from on-shell 
closed string modes.

Before analytic continuation, $V(0)/2$ is real and infinite, 
with $k_0\sim 0$. After analytic continuation, $V(0)/2$ becomes 
finite with an imaginary part. A comparison of (\ref{Vprop}),
(\ref{closedVz}) and (\ref{factor}) suggests that  
the analytic continuation moves the propagators on-shell, that is,
$Im~ V(0)/2$ is due to the sum over the on-shell poles. 
Since the divergence comes from large $n$, it further suggests 
%The potential we calculate obtains an imaginary part after 
%analytic continuation (which is tantamount to going on-shell) 
that most of $Im~ V(0)/2$ comes from massive on-mass-shell 
closed string modes. 
%Adding all the imaginary parts from each individual pole 
%from each asymptotic on-mass-shell closed string state gives 
%the imaginary part of the potential at $y=0$, $Im~V(0)$. 
For the massive closed 
string states, the poles get so dense for large $n$ that they are well 
represented by a branch cut. 
Our prescription for regularizing the divergence 
and isolating the imaginary part is equivalent to applying the 
$i\epsilon$ prescription of the analytic continuation to all 
asymptotically massive closed string modes and moving them on-shell.  
So $\Gamma = V_p Im~V(0)$ should be interpreted as the decay 
width of a non-BPS $Dp$-brane into on-mass-shell closed string modes. 

We may see this a little more clearly by separating the regularization
procedure into 2 steps: 
first go on-shell and then perform the analytic continuation.
Ignoring $\kpl$, we may restore the $k_0$ integral and rewrite 
(\ref{Vprop}) as
\baray
\frac{V(0)}{2} \sim \int^{\infty}dn\int dk_{\perp}^{9-p}\int dk_{0}
\delta(k^{2} - 4n/\ap) \theta(k_{0})
 \frac{A(n)}{\sqrt{\kp^2 + 4n/\ap}} \\ \nonumber
\to  \int dk_{\perp}^{9-p}\int_{0}^{\infty} dk_{0}
\frac{A(k^{2})}{k_{0}}
\earay
where the sum over $j$ is replaced by an integration over 
$n=\ap m^2/4$, with measure $A(n)$. Carrying out the $n$-integration 
with the delta function keeps only the on-mass-shell states.
Using the asymptotic form of $A(n)$ (\ref{an}), we find that 
$A(k^{2}) \sim k^{-11/2}~\exp(\sqrt{2\pi^{2}\ap k^2})$. 
This integral is divergent. After the analytic continuation of 
this expression, the integral becomes finite and obtains an 
imaginary part. Presumably, the inclusion of subleading 
terms of $A(n)$ will reproduce the correct $ Im~ V(0)$. 
%In summary, it is natural to interpret that, as we analytically continue 
%$V(0)$ so $Im~V(0)$ appears, the closed string modes move from off-shell
%to on-shell. That is, the decay width of a non-BPS-$Dp$-brane to 
%closed strings is given by $\Gamma = V_p Im V(0)$.

In Fig. 2, only bosonic closed string modes are involved.
They are the NS-NS and RR modes.   
In (\ref{closedVz}) (with $y=0$), we see that $\kp$ has a 
Gaussian distribution, with  $<\kp^2>= 2(9-p)/\ap s$.
From (\ref{zas}) and (\ref{Acontour}), we have 
$n \simeq 2 \pi^2/s^2$. Putting them together, we see that,
for massive states,
\baray
\frac{<\kp^2>}{m^2} \simeq \frac{9-p}{2 \pi \sqrt{2n}}
\earay
so the momenta of the massive closed string modes
transverse to the brane are negligible.
Conservation of momenta tangential to the 
brane implies that they are negligible too.
Since it is the large $n$ behavior that brings in the imaginary part, 
the decay is to very massive non-relativistic 
closed string modes, as pointed out in Ref \cite{liu,gir}. 
These massive closed string modes will then
decay to light (both bosonic and fermionic) closed string modes, 
which are expected to be relativistic. 

Apriori, the tachyon couples to other open string modes and 
the decay of a non BPS $Dp$-brane corresponds to the rolling 
of the tachyon \cite{sen1}, so one naively expects some 
energy to go to open string modes. However, no open string 
mode exists after the complete decay (and the disappearance) 
of the brane. This issue is resolved in Ref\cite{yi,sen3}: one 
expects the ends of an open string mode to have a flux tube 
($U(1)$) between them, so that the flux tube together with 
the open string form a closed string. As a result, only
closed string modes are produced. 

The estimate of $\Gamma$ is for the tachyon at or close to the 
top of the potential (at $\sqrt{2} \tau_p$). 
The time scale $t_T$ of the tachyon rolling 
is around $\sqrt{\ap}$.
In the compactified case where the world volume $V_p >> \ap^{p/2}$,
$t_T$ is comparable to or larger than the inverse of $\Gamma$.
In this case, the above estimate of $\Gamma$ should be valid.
For $t_T$ much smaller than $1/\Gamma$, we may expect the decay to 
start with the tachyon rolling, which goes to tachyon matter 
\cite{decay}, which then decays to relativistic closed strings. 
In this case, the above estimate may not be applicable.

$Dp$-branes in the bosonic string theory may also be considered : 
\baray
<Dp|~\Delta~|Dp> =  
\frac{1}{4\pi (8\pi^{2}\ap)^{13}} 
\int_0^{\infty}ds ~ e^{s} \prod_{m=1}^{\infty}
(1 - e^{-ms})^{-24}\\\nonumber 
\to \frac{1}{4\pi (8\pi^{2}\ap)^{13}}\int_0^{\infty} ds~ e^{s}~Z_B(s)
\earay
For $s \rightarrow 0$, 
$Z_B(s) \simeq \left({s}/{2\pi}\right)^{12}
\exp \left({4\pi^2}/{s} \right)$
so the degeneracy 
$D(n) \simeq 2^{-1/2} n^{-27/4} \exp ({4 \pi^2n}) $.
This yields for $s=2\pi/t$ and $n=v^2$ (ignoring powers of $t$ and $v$),
\baray
<Dp|~ \Delta~ |Dp> \sim \int_{0}^{\infty} dt 
dv \exp ({-2 \pi(v-t)^2/t + 2\pi t}) \nonumber
\earay
Completely analogous to (\ref{closedVn}), 
the last term in the exponent is identified as 
$-2 \pi \ap m_t^2 t=2 \pi t$ with the tachyon mass $\ap m_t^2=-1$. 
This is the origin of the apparent divergence in the integration 
over $t$.
Using the $s \to 0$ form of $Z_B(s)$ 
in $<Dp|~ \Delta~ |Dp>$, we obtain, after 
analytic continuation, the expression (\ref{imagVww}) for 
$2 Im <Dp|~ \Delta~ |Dp>$, with $p=25$ 
and $m_{tachyon}^2=-1/\ap$. Following the earlier analysis, we see 
that subleading terms of $Z_B(s)$ do not 
contribute to $Im ~<Dp|~\Delta~ |Dp>$.
 
It is suggestive to take $2 V_p Im ~V(y)$ (\ref{imaV}) to be 
the decay width of the $Dp$-$\Dbar p$-brane pair to closed strings,
since perturbatively, only closed strings are available after 
their annihilation. 
It will be important to understand this issue better.
For branes at an angle $\theta$, the branes recombine as 
$y \to 0$. Perturbatively, we expect the decay to release energy
to both open and closed strings. 

It is interesting to compare this result to that in quantum 
field theory. In quantum field theory, the propagator of 
a single field at the tree level has only $\delta$ finction
as its imaginary part.
Here the imaginary part appears classically due to the Hagedorn-like 
degeneracy of the closed string spectrum. However, 
this is a perturbative quantum effect in the open string channel. 

\bigskip
\bigskip

We thank Nick Jones, Finn Larsen, Hong Liu and Raul Rabadan for 
valuable discussions.
This material is based upon work supported by the National Science
Foundation under Grant No. PHY-0098631.

\section{Appendix :  Hardy-Ramanujan formula for the closed string 
level degeneracy}

Let $Z(s)=16 \Z(s)$ be given by:
\baray
\Z(s) = \prod_{m=1}^{\infty}\frac{(1+w^{m})^{D}}{(1-w^{m})^{D}} =
\sum_{0}^{\infty} \A(n)w^{n}
\earay
where $w = e^{-s}$. Using complex variables, we can write:
$A(n)=16 \A(n)$, where
\baray
\A(n)=\frac{1}{2\pi i} \oint dw \frac{\Z(w)}{w^{n+1}}
\earay
where the contour is a circle around the origin. To find the asymptotic
form of $\A(n)$, we write:
\be
\ln \Z(s)= D F(2s) - 2D F(s)
\ee
where $F(s)$ can be expressed in the Mellin representation:
\baray
F(s) = \sum_{n=1}^{\infty}\ln(1 - e^{-ns}) = -\frac{1}{2\pi i}
\int_{Re(z)=c}dz\Gamma(z)\zeta(1+z)\zeta(z)s^{-z},
\earay
where $c>1$, and $\zeta(z)$ is the Riemann zeta-function. The integrand
has a first-order pole at $z=1$ and a second-order pole at $z=0$.
Shifting the line of integration from $Re(z)= c >1$ to
$Re(z)= c^{\prime},-1< c^{\prime}<0,$ we arrive at:
\baray
F(s) = -\frac{\pi^{2}}{6s} - \frac{1}{2}\ln(\frac{s}{2\pi})-
\frac{1}{2\pi i}\int_{Re(z)
=c^{\prime}}dz\Gamma(z)\zeta(1+z)\zeta(z)s^{-z}.
\earay
Letting $z \to -z$ and using the identities:
\baray
&&\zeta(z)= 2^{z}\pi^{z-1}\sin(\frac{\pi z}{2})\Gamma(1-z)
\zeta(1-z),\nonumber \\
&&\Gamma(z)\Gamma(1-z)= \frac{\pi}{\sin(\pi z)},
\earay
one obtains :
\baray
F(s) = -\frac{\pi^{2}}{6s} - \frac{1}{2}\ln(\frac{s}{2\pi}) +
\frac{1}{2\pi i}\int_{Re(z)=c^{\prime \prime}}dz\Gamma(z)
\zeta(1+z)\zeta(z)(\frac{4\pi^{2}}{s})^{-z},
\earay
where $0< c^{\prime \prime} < 1$. Finally we move the path of
integration to the right side of the first-order pole at $z = 1$
and get the Hardy-Ramanujan formula:
\baray
F(s) = \frac{-\pi^{2}}{6s} - \frac{1}{2}\ln \left(\frac{s}{2\pi}\right)
+ \frac{s}{24} + F\left(\frac{4\pi^{2}}{s}\right)
\earay
In the limit ${s \to 0}, F(\frac{4\pi^{2}}{s}) = 0$,
and one obtains the asymptotic behavior of $\Z(s)$ as :
\baray
\Z(s) \simeq \left(\frac{s}{4\pi}\right)^{\frac{D}{2}} 
e^{\frac{D\pi^{2}}{4s}}
\earay
Using this result and carrying out a saddle point evaluation of the
contour integral, we get:
\baray
\label{a2}
\A(n)\simeq \left(\frac{D}{64}\right)^{(D+1)/4}n^{-(D+3)/4}
e^{\sqrt{D\pi^{2}n}} \left( 1 - \frac{10}{\pi\sqrt{n}}+ ... \right)
\earay
For general $s$,
\baray
\ln \Z(s)= \frac{D\pi^{2}}{4s} + \frac{D}{2}\ln \left(\frac{s}{4\pi}\right)
 -De^{-2\pi^{2}/s}+ \frac{D}{2}e^{-4\pi^{2}/s}-\frac{D}{3}
e^{-6\pi^{2}/s}+\frac{5D}{4}e^{-8\pi^{2}/s} + .....
\earay
This leads to the following asymptotic expansion for $\Z(s)$:
\baray
\label{a3}
\Z(s)= \left(\frac{s}{4\pi}\right)^{\frac{D}{2}}
e^{\frac{D\pi^{2}}{4s}} \left[ 1 - \right. De^{-2\pi^{2}/s} +
\left( \frac{D^{2}}{2}+\frac{D}{2} \right) e^{-4\pi^{2}/s}-
\left( \frac{D^{3}}{6}+\frac{D^{2}}{2} 
+\frac{4D}{3} \right) e^{-6\pi^{2}/s} \\\nonumber
 + \left( \frac{D^{4}}{24}
+\frac{D^{3}}{4}+\frac{35D^{2}}{24}+\frac{5D}{4} \right)
e^{-8\pi^{2}/s} - ..... \left. \right]
\earay

A similar (and simpler) analysis can be carried out for the bosonic string.

\end{document}